# Graphene ribbons with suspended masses as transducers in ultra-small nanoelectromechanical accelerometers

Xuge Fan[1]★, Fredrik Forsberg[2], Anderson D. Smith[3], Stephan Schröder[1, 4],

Stefan Wagner[5], Henrik Rödjegård[4], Andreas C. Fischer[1, 6], Mikael Östling[3],

Max C. Lemme[3, 5, 7]★, Frank Niklaus[1]★

[1]Department of Micro and Nanosystems, School of Electrical Engineering and Computer Science, KTH Royal Institute of Technology, SE-10044 Stockholm, Sweden.

[2]Scania Technical Centre, 15148 Södertälje, Sweden.

[3]Department of Integrated Devices and Circuits, School of Electrical Engineering and Computer Science, KTH Royal Institute of Technology, SE-164 40 Kista, Sweden.

[4]Senseair AB, 820 60 Delsbo, Sweden.

[5]AMO GmbH, Otto-Blumenthal-Str. 25, 52074 Aachen, Germany.

[6]Silex Microsystems AB, 175 26 Järfälla, Sweden.

[7]Chair of Electronic Devices, Faculty of Electrical Engineering and Information Technology, RWTH Aachen University, Otto-Blumenthal-Str. 25, 52074 Aachen, Germany.

★email: xuge@eecs.kth.se, lemme@amo.de, frank.niklaus@eecs.kth.se

**Nanoelectromechanical system (NEMS) sensors and actuators could be of use in the development of next generation mobile, wearable, and implantable devices. However, these NEMS devices require transducers that are ultra-small, sensitive and can be fabricated at low cost. Here, we show that suspended double-layer graphene ribbons with attached silicon proof masses can be used as combined spring-mass and piezoresistive transducers. The transducers, which are realized using processes that are compatible with large-scale semiconductor manufacturing technologies, can yield NEMS accelerometers that occupy at least two orders of magnitude smaller die area than conventional state-of-the-art silicon accelerometers. With our devices, we also extract the Young's modulus values of double-layer graphene and show that the graphene ribbons have significant built-in stresses.**

Ultra-small nanoelectromechanical system (NEMS) accelerometers have a range of potential applications, including the Internet of Things (IoT) [1], wearable electronics for monitoring activity levels[2] and patient recovery[3], and implantable systems for monitoring heart failure[4]. However, creating such NEMS devices requires electromechanical transducers that can be aggressively down-scaled to device dimensions of a few tens of square micrometres, while retaining high device sensitivity. Graphene is a promising material for electromechanical transducers because of its atom-layer thinness, and its unique electrical and mechanical properties[5,6]. NEMS transducers could, in particular, be created by using suspended graphene ribbons with attached heavy proof masses and integrated piezoresistive transduction. However, while suspended graphene membranes and beams have been used to characterize the material properties of graphene[7-13], demonstrations of their practical application in NEMS have so far been limited to basic structures such as resonating beams[10,11] and pressure sensor membranes[11-13].

In this Article, we show that ultra-small spring-mass systems with piezoresistive transducers can be created by suspending silicon proof masses on double-layer graphene ribbons. Characterization of the mechanical and electromechanical properties of the suspended graphene ribbons, as well as the dynamic properties of the spring-mass systems, shows that they are useful transducers for NEMS accelerometers with dramatically reduced dimensions and increased performance. With our system, we also extract the Young's modulus values of double-layer graphene fabricated by layer stacking. Furthermore, we find that the graphene ribbons have significant built-in stresses, which have a tangible influence on the static and dynamic characteristics of the devices, consistent with work on the effects of built-in stress in graphene ribbons and membranes[7,10,14,15]. Our graphene NEMS transducers are compatible with large-scale semiconductor fabrication technologies[16] and could be used to create ultra-miniaturized NEMS accelerometers, gyroscopes and microphones, for potential applications in biomedical implants, nanoscale robotics, vehicle safety systems, consumer electronics, wearable electronics and the IoT.

**Device fabrication**

We fabricated suspended graphene ribbons with attached silicon proof masses using two layers of chemical vapour deposited[17] (CVD) graphene that were transferred to an oxidized silicon-on-insulator (SOI) wafer with etched trenches in the silicon device layer of the SOI wafer (Fig.1a and 1b; see Methods and Supplementary Sections S1-S4 for details of device fabrication and structural evaluation) and etched cavities in the silicon handle substrate below the trenches (Fig.1b1-1b2). Next, the graphene was patterned to form ribbons and thus, the proof masses defined in the silicon device layer were sandwiched between the graphene ribbons and the buried $SiO_2$ (BOX) layer of the SOI wafer (Fig.1b3). The BOX layer in the areas below the trenches in the silicon device layer was sacrificially removed by dry plasma etching followed by vapour hydrogen fluoride (HF) etching to carefully release the proof masses and suspend them on the graphene ribbons (Fig.1b4). SEM images of fabricated devices are shown in Fig.1 c1-c4. To electrically characterize the finalized devices, they were placed in a ceramic package and wire-bonded (Fig.1c5-c7). Devices with different dimensions were fabricated and evaluated (for detailed device dimensions see Supplementary Section S5, Table S1). The trench width, i.e. the dimension defining the length of the freely suspended graphene ribbons range from 2-4 µm. The silicon proof masses of all resulting devices are 16.4 µm thick and have a quadratic shape with side lengths ranging from 10-50 µm. Thus, these proof masses are three to seven orders of magnitude heavier than the thin coatings deposited on suspended graphene that have been previously reported in literature[18-20] (Supplementary Section S6, Table S2).

For our devices we used double-layer graphene, which resulted in fabrication yields of the suspended ribbons with attached proof masses of well above 50%. We also attempted to use single-layer graphene for realizing suspended ribbons with attached proof masses, however the resulting fabrication yield was on the order of 1% because the structures were extremely fragile and it was difficult to handle the devices without breaking them. While suspended mechanically exfoliated graphene has extremely high intrinsic strength[6] (resistance to deformation), the fracture toughness (resistance to fracture) is a more relevant parameter here, and recent reports suggest that the fracture toughness of polycrystalline single-layer CVD graphene is relatively moderate[21, 22]. Our experimental results confirm that, for a suspended structure made of CVD graphene, the addition of a second CVD graphene layer on top of a first CVD graphene layer disproportionally increases the fracture toughness of the resulting structure, consistent with literature reports of increased overall mechanical resilience of suspended double-layer CVD graphene membranes as compared to single-layer CVD graphene membranes [23].

## Acceleration measurements and performance analysis

We evaluated the viability of suspended graphene ribbons with an attached proof mass for use as a NEMS accelerometer and our characterization revealed how exposing the proof mass to acceleration forces results in resistance changes of the graphene strain gauge. The principle of operation of this type of accelerometer is based on the displacement of the proof mass caused by acceleration forces acting on the proof mass. The resulting strain that builds up in the suspended graphene ribbons causes resistance changes in the graphene ribbons because of the piezoresistivity in graphene. Larger strain in the graphene ribbons, for example caused by a larger applied acceleration, results in a larger resistance change in the graphene ribbons. In our accelerometer designs, the suspended graphene ribbons simultaneously form the springs of the spring-mass system and the piezoresistive transducer elements. The die area occupied by these functional elements is more than two orders of magnitude smaller than the one occupied by the equivalent functional elements of conventional piezoresistive and capacitive accelerometers (Supplementary Section S7, Table S3). For the characterization of the conversion between vibration and output signal of our devices we used an air-bearing shaker with a built-in reference accelerometer. In all experiments, the devices were placed with the sensitive axis in the direction of the earth gravitation, providing a 1 g acceleration bias. Careful electrostatic and magnetic shielding of our devices was used to minimize the pickup of parasitic noise signals (Methods and Supplementary Section S8).

The spectrums of the amplified output voltages (amplification factor of ~500) of device 1 and a reference device while exposed to a nominal acceleration of 1 g at a frequency of 160 Hz (as measured by the built-in reference accelerometer) are shown in Fig.2a and d, respectively (Supplementary Sections S9-S13 for more devices and control measurements). The reference device consisted of a graphene strip similar to the strips in the accelerometer devices, but without etched trenches in the underlying substrate surface. The output signal of device 1 correlated with the applied acceleration (red peaks in Fig.2a), while no apparent signal was present in the output of the reference device (black curve in Fig.2d). This indicates that it is the acceleration-induced displacement of the proof mass that caused the measured resistance change in the graphene strain gauge, and that the resistance change is not significantly influenced by parasitic electromagnetic signals induced in the graphene (Supplementary Section S9). As expected, we found that increased acceleration yielded a higher output voltage signal (Fig.2b). We also measured the noise spectral density of device 1 and two additional devices (Supplementary Section S10), illustrating that the 1/f-noise in these devices is comparably moderate, which is in agreement with the relatively low 1/f-noise of resistive graphene patches reported in literature[24-26]. When exposed to an acceleration of 1 g at a frequency of 160 Hz, the resistance change ($\Delta R$) of device 1 was ~ 41 m$\Omega$, and the relative resistance change ($\Delta R/R$) of device 1 was ~ 0.0028 % (Fig.2b and c). The noise density of device 1 is estimated to be on the order of 50 µg/$\sqrt{Hz}$ at 100 Hz, and is limited by 1/f-noise. This corresponds to a resolution of the order of 0.68 mg in the frequency range from 16 to 100 Hz. While this value is lower than the resolution of some silicon accelerometers reported in literature, for example 0.013 mg[27] and 0.02 mg[28] in the frequency range from DC to 100 Hz, with noise densities of 4.53 µg/$\sqrt{Hz}$ and 20 µg/$\sqrt{Hz}$, respectively, the silicon piezoresistive accelerometers have proof masses that are more than three orders of magnitude larger than the proof masses of our devices (Supplementary Section S7). From the perspective of the dominating electronic noise in our devices, the resolution can be improved by using graphene of high quality with low 1/f-noise, and by optimizing the measurement circuit.

To explore the impact of device design variations on the resulting output signal, we compared the output signal of two similar devices, device 2 and device 3 (Fig. 2g), with identical masses (30 µm × 30 µm × 16.4 µm) and lengths of the suspended graphene ribbons (2 × 3 µm), but different widths of the ribbons (13 µm for device 2 and 24 µm for device 3). In these experiments device 2 yielded higher output voltages (Fig. 2e) and resistance changes (Fig. 2f) than device 3 when exposed to defined accelerations. As expected, these results suggest that reducing the width of the suspended graphene ribbons and thereby increasing the ensuing strain results in a larger output signal under identical conditions. We measured the relative resistance change ($\Delta R/R$) at an applied acceleration of 1 g at a frequency of 160 Hz of a total of eleven different devices, yielding comparable relative resistance changes for these devices (Supplementary Section S7). To evaluate the functionality of a device that has the proof mass attached with an offset relative to the centre line of the graphene ribbons, we have realized and measured a device with an offset of the proof mass attachment of approximately 2 µm (device 4, Fig. 2g) and obtained a signal response that was of the same order as the response of devices with centred proof masses. No obvious differences in device functionality were observed (Fig. 2h and i).

To verify that parasitic effects such as influences from humidity or gas flow in the vicinity of the graphene ribbons were not causing the signal response of our devices, we placed a device (device 5) inside a ceramic package with an actively pumped vacuum of ~$10^{-5}$ bar and exposed it to an acceleration of 1 g at a frequency of 160 Hz. We found that the signal response of the device was not significantly changed as compared to operating the device at atmospheric pressure (Supplementary Section S11). To evaluate the reproducibility of the measured output signal of the same devices, we measured the same device (device 6) at different times while exposing it to an acceleration of 1 g and observed good reproducibility of the output signal of > 95% (Supplementary Section S12). To evaluate the reproducibility of the properties of different devices with comparable designs, two devices with comparable designs, devices 1 and 7, were measured while exposed to an identical nominal acceleration of 1 g at a frequency of 160 Hz, showing comparable output voltages (Fig. 2a and Supplementary Section S13). While the signal responses of the two devices were similar, the existing differences may be attributed to variations in the graphene and device dimensions caused by the fabrication process, including the presence of a defect in one of the graphene ribbons of device 7 (Supplementary Section S13).

# Mechanical properties

The utilization of suspended graphene ribbons with attached silicon proof masses as NEMS transducers requires understanding of their physical and mechanical properties. We characterized the static and dynamics frequency characteristics of our devices by atomic force microscopy (AFM) and laser Doppler vibrometry (LDV), respectively (Methods). To deeply explore the mechanical behaviour of our devices, we developed a finite element analysis (FEA) description (Supplementary Section S14) and a detailed analytic description (Supplementary Section S15) of the device structures. We used these models to extract the Young's modulus, built-in stress, and spring properties of the suspended double-layer graphene ribbons, and to analyse the frequency behaviour of the spring-mass systems. In our analytic description of the device structure, we approximated the two suspended graphene ribbons with attached proof mass by a single doubly-clamped ribbon with a centre-point load (Supplementary Section S15). For large deflections of the ribbon compared to the thickness of the ribbon, the deflection at the centre of the ribbon caused by a centre-point load is described by

$$F = 16 \left[\frac{EWH^3}{L^3}\right] Z + 8 \left[\frac{EWH}{L^3}\right] Z^3 + 4 \left[\frac{T}{L}\right] Z \qquad (1)$$

where F is the load applied at the centre of the ribbon, Z the resulting deflection of the ribbon at its centre, E the Young's modulus of the graphene, W the width of the ribbon, H the thickness of the ribbon, L the total length of the ribbon, and T the built-in tension force of the ribbon (Supplementary Sections S15 and S16). The average residual built-in stress ($\sigma_0$) in a doubly-clamped ribbon can be approximated by $\sigma_0 = \frac{T}{WH}$ (Supplementary Section S15). A comparison with our FEA simulation results demonstrates that equation (1) correctly describes the displacement of a suspended graphene ribbon with an accuracy of > 97% over a wide range of deflections, which is significantly more accurate than the analytic models previously reported in literature[9,29-32] (Supplementary Section S15). We measured the force-displacement curves on four different devices (devices 8 to 11, see Supplementary Section S5 for device dimensions) using AFM indentation at the centre of the suspended proof mass (Fig. 3a and Methods), revealing a near perfect match of the measurement results to the force-displacement curves predicted by both, equation (1) and our FEA model. Fitting of the force-displacement data of devices 8, 9, 10 and 11 to equation (1) yielded Young's modulus values of the suspended double-layer graphene of 0.23 TPa, 0.21 TPa, 0.25 TPa and 0.21 TPa, and built-in stresses in the graphene ribbons of 318 MPa, 351 MPa, 396 MPa and 316 MPa, respectively (Fig. 3b and d). For comparison, a fit of the AFM measurement data of devices 8 and 9 to equation (1) for negligible built-in stress values in the graphene ribbons ($\sigma_0 \sim 0$ Pa) yielded Young's modulus values of 0.32 TPa and 0.35 TPa, respectively, but resulted in poor curve fitting (Fig. 3c). The AFM indentation force used in these force-deflection measurements cover a range from 10 to 1000 nN for devices 8 and 9 and from 10 to 200 nN for devices 10 and 11, resulting in proof mass deflections of up to 374 nm for the highest indentation force of 1000 nN (largest force available in our AFM tool). The large measured force-deflection range in our experiments, together with the near perfect match to theory without any outlier data points, is strong evidence that our extracted Young's modulus values and built-in stresses are reliable. To further evaluate the quality of the curve fit for device 8, we removed individual data points, one at a time, which affected the fitted curve with less than 10%.

To evaluate the torsional robustness of our devices, we conducted force-displacement measurements using AFM indentation at positions on the proof masses that are offset in relation to the centre line of the graphene ribbons (Supplementary Section S17). For a force of 1000 nN with an indentation offset of ~4 μm, the proof mass deflection was up to 1400 nm (device 8, Supplementary Section S17). This force corresponds to an acceleration of > 6000 g acting on the $1.52 \times 10^{-8}$ g proof mass of device 8. No device failure was observed in any of our AFM indentation experiments, confirming that the devices are mechanically robust and that the graphene ribbons can survive very large deflections. To the best of our knowledge, the forces we have introduced on our suspended doubly-clamped graphene ribbons by AFM indentation and the resulting deflections and uniaxial strain levels in the graphene ribbons, are of the same order[9] or larger[29,30,33,34] than the ones introduced by AFM indentation on suspended doubly-clamped graphene beams or ribbons reported in literature (Supplementary Section S17, Table S8). Some of the previously reported fully-clamped graphene membranes[6] also can survive very large AFM indentation forces (e.g. 1200 nN), resulting in large bi-axial strain levels in the fully-clamped graphene membranes (Supplementary Section S17, Table S8), which however is a different type of strain compared to the uniaxial strain in our devices. In our acceleration measurements described above, we have exposed graphene devices to a maximum acceleration of 2 g, which was a limit set by the specifications of the shaker in combination with the mass of the shielding box. We did not observe device damage in these experiments, suggesting that the useful dynamic range of our devices is at least ± 2 g, but likely significantly larger than this value, i.e. hundreds or thousands of g. When our graphene devices were exposed to excessive shocks during handling we have in some cases observed that the devices were damaged. The dominant failure mode in these cases was stiction of the proof mass to the sidewall of the trench (Supplementary Section S18).

To explore the resonant properties and frequency modes of our devices and to obtain an independent estimate of the built-in stresses in the graphene ribbons, we measured the resonance frequencies of the spring-mass systems of our devices using LDV (Methods and Supplementary Section S19) and compared the measurement results to the modal analysis based on our FEA model and to a standard linear analytic model of our devices. The modal analysis predicted that, of the six degrees of freedom of the proof mass (three translational and three rotational rigid body motions), the resonance mode perpendicular to the ribbon surface along the z-axis (Z-mode) is most easily excited (Fig. 4a and Supplementary Section S14). The modal analysis also showed that possible manufacturing-induced misalignments between the proof mass and the graphene ribbons that are of the order of up to a few micrometres, have no significant impact on the resulting resonance frequencies (Supplementary Section S14). The FEA results depicting the dominant Z-mode movement of the proof mass is illustrated in Supplementary Video 1. The LDV measurements showed that the most probable Z-mode resonance frequencies ($f_0$) of the spring-mass systems of devices 12, 13, 14 and 15 are 14.9 kHz, 16.2 kHz, 24.2 kHz and 27.2 kHz, respectively (Fig.4b and c, and Supplementary Section S19, Figure S20 and Tables S9-S12). A video taken by LDV of the resonant Z-mode movement of the proof mass of device 14 is available as Supplementary Video 2. The resonance frequency limits the potential maximum bandwidth of a device, which for devices 12 to 15 can be estimated to be from DC to 14.9 kHz, from DC to 16.2 kHz, from DC to 24.2 kHz, and from DC to 27.2 kHz, respectively. These are bandwidths that are of the same order as the bandwidths reported for typical silicon accelerometers[27, 28, 35].

Based on the measured resonance frequencies and the dimensions of devices 12, 13, 14 and 15 we used our FEA model to extract estimated built-in stresses in the graphene ribbons, arriving at values of 228 MPa, 441 MPa, 231 MPa and 291 MPa, respectively (Fig. 4d), which are of the same order as the built-in stresses extracted by the AFM indentation measurements of devices 8, 9, 10 and 11 (Fig. 3d). These built-in stresses are consistent with reported typical built-in stresses of fully-clamped graphene membranes ($\sim 10^2$ to $10^3$ MPa), but somewhat larger than reported typical built-in stresses of doubly-clamped ribbons measured by AFM indentation (~ $10^1$ MPa) (Supplementary Section S20, Table S14). Built-in stresses in suspended graphene are affected by multiple factors, including the design features of the sample, the type of graphene growth, and the graphene transfer process[23]. A factor contributing to the built-in stresses in our devices may be pre-straining of the suspended graphene ribbons as a result of van der Waals attraction between the suspended graphene ribbons and the $SiO_2$ substrate surfaces at the edges of the trenches[6, 7, 10, 36, 37], which is consistent with geometrical considerations of the situation at the graphene anchor positions (Supplementary Section S21) and with the measured static deflections of the proof masses in our devices using white light interferometry (Supplementary Section S3). The influence of the weight of the proof mass on the extracted built-in stresses is very small and can be ignored. For the measured built-in stresses in our suspended graphene ribbons (~ 230 to 440 MPa) and the small acceleration levels introduced by the vibration excitation through the shaker ($\leq$ 1 g), equation (1) is dominated by the built-in tension T. In this case the system can be approximated as linear and we can use Newton's second law and the standard linear accelerometer model based on a spring-damper-mass system to arrive at the mechanical transfer function[38]

$$H(s) = \frac{1}{s^2 + \frac{D}{M}s + \frac{K}{M}} = \frac{1}{s^2 + \frac{2\pi f_0}{Q}s + (2\pi f_0)^2} \quad (2)$$

Where s is the complex variable, Q is the quality factor, D is the damping factor, M is the mass and K is the effective spring constant of the system. Equation (2) yields $K = (2\pi f_0)^2 M$. For this case, and assuming that the Young's modulus is in a range between 0.1 to 1 TPa, the resonance frequency of the system is mainly impacted by the built-in stress and not by the Young's modulus, which was independently confirmed by our FEA results. The extracted effective spring constants K of devices 12, 13, 14 and 15 are 0.542 N/m, 0.618 N/m, 1.431 N/m and 1.806 N/m, respectively, consistent with typical values of linear spring constants of comparable graphene ribbons reported in literature[10,29,30] (Supplementary Section S22, Table

S15). It should be noted that for large deflections of the graphene ribbons (i.e. for large acceleration forces acting on a proof mass) and/or for ribbons with low built-in stress, the system does not behave in a linear way. In these cases, a non-linear model has to be used to describe the system.

**Electro-mechanical properties**

To elucidate the electro-mechanical response of our devices we used both, the FEA model and equation (1) to estimate the expected deflections of the suspended graphene ribbons, assuming the measured average Young's modulus of 0.22 TPa of the graphene and an estimated built-in stress of 250 MPa of the graphene ribbons. For these parameters and acceleration levels of the order of 0.1 to 1 g along the z-axis, the calculated centre point displacements of the proof masses and the suspended graphene ribbons in our devices are below 1 nm, following a linear relationship with acceleration. While the measured resistance response of our devices in dependence of the acceleration has a positive correlation (Fig.2), the theoretic deflection of the graphene ribbons predicted by our simplified analytic model cannot completely account for the measured resistance responses resulting from the piezoresistivity of graphene. For intrinsic resistance responses ($\Delta R_{SG}/R_{SG}$, $R_{SG}$ refers here to the combined resistance of the suspended graphene ribbons, while $\Delta R_{SG}$ refers to corresponding resistance change) that are of the order of 0.02 % as extracted in our measurements (e.g. device 1), the deflections of the suspended graphene ribbons should be on the order of 10 nm if realistic gauge factors of the order of 10 are assumed[12, 39-42]. It should be noted that the reported gauge factors of graphene strain gauges span a wide range and are still being debated in the research community (Supplementary Section S23, Table S16). According to our simplified device model and assuming a built-in stress of 250 MPa in the graphene ribbons, to obtain proof mass and ribbon deflections of the order of 10 nm, the effective acceleration levels should be of the order of 20 to 30 g. Alternatively, to obtain deflections of 10 nm at effective acceleration levels of 1 g, the built-in stress in the graphene ribbons should be of the order of 1 MPa.

To further explore the actual proof mass deflections of our devices, we used LDV to measure the proof mass deflections of devices at applied accelerations of 0.5 g and 1 g and at different fixed frequencies. For the same device, different proof mass deflections were measured for identical accelerations but different excitation frequencies (Fig. 4e, Supplementary Section S19, Table S13 and Supplementary Video 3). For example, in device 12 the averaged measured displacements of the proof mass at applied accelerations of 0.5 g and 1g at a frequency of 160 Hz were about $11 \pm 2$ nm and $22 \pm 3$ nm, respectively, while the measured proof mass displacements at applied accelerations of 0.5 g and 1g at a frequency of 9.725 kHz were about $0.75 \pm 0.25$ nm and $1.25 \pm 0.25$ nm, respectively. Using equation (S7) in Supplementary Section S15, proof mass displacements of 0.75 nm, 1.25 nm, 11 nm and 22 nm in device 12 correspond to strains in the graphene ribbons of approximately $1.8 \times 10^{-8}$, $4.9 \times 10^{-8}$, $3.8 \times 10^{-6}$ and $1.5 \times 10^{-5}$, respectively. In device 13, with identical proof mass size and graphene ribbon length but more narrow ribbons as compared to device 12, the measured proof mass displacements at applied accelerations of 0.5 g and 1g and a frequency of 160 Hz were about $17 \pm 2$ nm and $31 \pm 4$ nm, respectively, and about $10 \pm 1$ nm at an applied acceleration of 1g and a frequency of 16 kHz. Proof mass displacements of 10 nm, 17 nm and 31 nm in device 13 correspond to strains in the graphene ribbons of approximately $3.1 \times 10^{-6}$, $9 \times 10^{-6}$ and $3 \times 10^{-5}$, respectively. The larger measured proof mass deflections in device 13 as compared to device 12 are in line with expectations due to the wider ribbons in device 12. Our LDV measurements indicate that the effective acceleration acting on the proof mass may be significantly higher than the 1 g acceleration measured by the reference accelerometer of the shaker. Such an effect can be caused by vibration modes of parts that do not belong to the spring-mass system of the device, including parts of the package, measurement set-up or read-out circuitry. If an external part is going into resonance it can amplify the applied vibration, increasing the amplitude and thus, significantly increasing the effective acceleration to which the proof mass is exposed. Such resonance effects can potentially be reduced by reducing the size and mass of the electromagnetic shielding box and the measurement circuitry, for example by integrating the graphene device in a chip-scale package with integrated CMOS read-out circuits. Another possible factor contributing to the measured higher than expected resistance changes in our devices could be effects caused by crumpling of the suspended graphene ribbons. It was reported that suspended graphene is inevitably crumpled in the out-of-plane direction, both, due to dynamic out-of-plane flexural phonons, and due to static wrinkling that may be caused by uneven stress at the boundary of graphene produced during device fabrication[43]. Such crumbling can impact the mechanical and electrical behaviour of suspended graphene, especially for small deflection and strain levels in the suspended graphene[43]. In addition, it is also possible that interactions between the graphene ribbons and the $SiO_2$ surface at the trench edges and/or related delamination effects contribute to the measured resistance changes in the devices, although our analysis suggests that such effects are less likely to cause significant resistance changes in the graphene ribbons (Supplementary Section S24).

## Conclusions

We have reported suspended graphene ribbons with attached silicon masses for use as transducers in NEMS devices such as accelerometers and resonators. We used the system to create NEMS accelerometer structures that occupy die areas that are at least two orders of magnitude smaller than the die areas occupied by the most compact state-of-the-art silicon accelerometers (Supplementary Section S7, Table S3). For example, for device 1 in our study (Fig. 2a), the die area occupied by the functional elements – that is, the proof mass, graphene transducer, anchor frame and electrical contacts – is about 80 μm x 60 μm. This can be compared to die areas occupied by the respective elements of typical state-of-the-art piezoresistive accelerometers that are on the order of 2900 μm x 1000 μm[44]. There is also potential to reduce further the dimensions of our graphene devices by minimizing the electrical contact areas and by using state-of-the-art device packaging strategies[45].

In order to compare the relative resistance changes in piezoresistive accelerometers with dissimilar proof mass sizes, the normalized relative resistance change per proof mass volume is a suitable figure of merit. The proof masses of our devices are at least three orders of magnitude smaller than the proof masses of commonly reported piezoresistive silicon accelerometers (Supplementary Section S7). In device 1, for example, with the assumption that the equivalent acceleration acting on the proof mass is 30 g, the relative resistance change ($\Delta R/R$) per proof mass volume would be about one order of magnitude larger than for previously reported piezoresistive accelerometers[44,46-48] (Supplementary Section S7, Table S4 and Figure S6).

Based on the analytic and FEA descriptions of our devices, the design parameters that can further improve device performance include increasing the mass of the proof mass, reducing the widths of the suspended graphene ribbons and reducing the built-in stress in the graphene ribbons. In addition, increasing the length of the suspended graphene ribbons can increase the output signal to some extent by increasing the absolute resistance change ($\Delta R_{SG}$) due to the increase of the combined resistance of the suspended graphene ribbons ($R_{SG}$). It should be noted that the length of the graphene ribbons is expected to have a negligible effect on the resulting change of the strain of the suspended graphene ribbons (Supplementary Section S15). However, there are various design trade-offs that place constrains on these parameters, including the desired device dimensions, the fabrication yield and the device robustness. Furthermore, the implementation of ribbons made of other 2D materials, such as $MoS_2$[49], with significantly higher piezoresistive gauge factors than those of graphene would be an interesting approach to improve the performance of our devices. While a graphene-based shock detector has been proposed[18], which uses capacitive transduction with a noise-equivalent signal pulse when exposed to shocks of 1000 to 3000 g, no functional graphene-based NEMS accelerometer has been reported to date.

From our experiments, we extracted an average Young's modulus for double-layer graphene of 0.22 TPa and non-negligible built-in stresses of the order of 230 to 440 MPa in the suspended graphene ribbons. Although the extracted Young's modulus of double-layer CVD graphene is lower than the commonly reported[6,50] value of 1 TPa for single-layer graphene, our results are consistent with reports of reduced Young's modulus values of single-layer CVD graphene (~0.164 TPa[23], ~0.3 TPa[43]), few-layer graphene made of chemical reduced graphene oxide (~0.25 TPa[9]). The lower Young's modulus values in double-layer graphene ribbons are presumably due to competing bending and shear forces in stiffened structures[9,51]. For instance, for stiffer structures such as multilayer graphene, bending deflections may be negligible and shear strains may play a dominant role[9]. Furthermore, grain boundaries as well as ripples that are inherent in the structure of the CVD graphene[23], different types and density of defects[52], and interlayer sliding-induced energy dissipation between graphene layers resulting from relatively weak interlayer adhesion between graphene layers[53,54] can affect the stiffness of graphene and cause a lower Young's modulus in double-layer graphene compared with single-layer graphene. It has also been reported that crumbling of suspended graphene due to static wrinkling can result in reduced Young's modulus values for relatively small deflection and strain levels applied to suspended graphene[23,43,55]. However, this effect gradually reduces for larger applied deflection and strain levels and we do not expect that it has a significant influence on our results, as the force-deflection measurements we used for extracting our Young's modulus values involve especially large deflection and strain levels of the graphene ribbons. For reference, a detailed comparison of Young's modulus values of different graphene-based materials is provided in Supplementary Section S25, Table S17.

# Methods

## Device fabrication

**Substrate preparation:** Devices were fabricated from a SOI substrate in which the silicon device layer is 15 μm thick, the BOX layer is 2 μm thick and the handle substrate is 400 μm thick (Fig.1a). First, the SOI wafer was thermally oxidized to grow a 1.4 μm thick layer of $SiO_2$ on both the front and the backside of the wafer. A photoresist layer was spin-coated on the $SiO_2$ surface of the silicon device layer and patterned for defining the metal electrodes. The pattern was transferred into the 1.4 μm thick $SiO_2$ layer by etching 300 nm deep cavities using reactive ion etching (RIE). Next, the cavities were filled with a 50 nm thick layer of titanium (Ti) followed by a 270 nm thick layer of gold (Au) using evaporation. The photoresist layer was removed in a lift-off process by wet etching, leaving the patterned Au electrodes protrude by ~20 nm above the $SiO_2$ surface. A new photoresist layer was spin-coated on the $SiO_2$ surface and lithographically patterned for defining the trenches surrounding the proof masses. RIE was used to etch through the 1.4 μm thick $SiO_2$ layer and DRIE was used to etch through the 15 μm thick silicon device layer (Fig.1b1). After the DRIE, photoresist residues were removed by $O_2$ plasma. Next, a photoresist layer was spin-coated and patterned on the $SiO_2$ surface of the backside of the SOI wafer, defining squares with dimensions of 150 μm × 150 μm placed in the same areas that define the proof masses in the silicon device layer. The 1.4 μm thick $SiO_2$ layer was etched by RIE. Next, the handle substrate was etched by DRIE until reaching the BOX, using both the photoresist and the $SiO_2$ as masking layers. Photoresist residues were removed by $O_2$ plasma, which finalized the pre-processing of the SOI device substrate (Fig.1b2). The device substrate was then diced in 8 mm × 8 mm large chips, each containing 64 devices.

**Graphene transfer and patterning:** Commercially available CVD single-layer graphene on copper foil (Graphenea, Spain) was used in this work. Double-layer graphene was obtained by transferring a single-layer graphene to another single-layer graphene on a copper foil. Therefore, a poly (methyl methacrylate) (PMMA) solution (AR-P 649.04, ALLRESIST, Germany) was spin-coated on the front-side of the first graphene/copper foils at 500 rpm for 5 seconds and at 1800 rpm for 30 seconds and then baked for 5 minutes at 85°C on a hot plate to evaporate the solvent and cure the PMMA, resulting in a film thickness of ~200 nm. Carbon residues on the backside of the copper foil were removed using $O_2$ plasma etching at low power (50-80 W). In order to release the graphene/PMMA stack from the copper, the foil was placed onto the surface of an iron chloride ($FeCl_3$) solution with the graphene side facing away from the liquid, resulting in wet etching of the copper. After 2 hours, the PMMA/graphene stack without copper floating on the $FeCl_3$ solution was transferred onto the surface of deionized (DI) water, then diluted HCl solution and, back to DI water for cleaning, removing the iron (III) residues and removing chloride residues, respectively. A silicon wafer was used for handling and picking up the PMMA/graphene stack from the liquids. During the etching and cleaning processes, it is important to keep the PMMA/graphene stack floating on the surface of the liquids and keep the graphene side on top, in order to make sure that the PMMA covering the graphene is not wetted by the etch solution. A second graphene on copper foil was used and the PMMA/graphene stack floating on the DI water was transferred to the second graphene on copper foil and subsequently put on a hotplate at 45°C to increase the adhesion between the two graphene layers. Carbon residues on the backside of the copper foil were removed using $O_2$ plasma. Again, the same processes were performed to remove the copper substrate from the double-layer graphene and transfer the final PMMA/double-layer graphene stack to the pre-processed SOI device substrate. The device substrate was then baked for 10 minutes at 45℃ in order to dry it and to increase the bond strength between the double-layer graphene and the SiO$_2$ substrate surface. Next, the device substrate was placed into acetone for 24 hours to remove the PMMA and subsequently into isoproponal for 5 minutes to remove acetone residues. A nitrogen gun was used to gently dry the device substrate, followed by baking at 45℃ for 10 minutes on a hot plate. A detailed schematic diagram of the graphene transfer process for realizing double-layer graphene and subsequently transferring the double-layer graphene to the SOI device substrate is presented in Supplementary Section S1. After graphene transfer, a photoresist layer was spin-coated on the graphene at 1000 rpm for 5 seconds and 4000 rpm for 60 seconds and then baked for 30-60 seconds at 90℃ on a hotplate. Optical lithography and photoresist development was done using a standard developer for 15 seconds and DI water for 10 seconds for rinsing, and then the SOI substrate was dried in air. Next, the graphene was etched by O$_2$ plasma at 50 W for 120 seconds to define the graphene ribbons. Finally, in order to remove the photoresist residues, the device substrate was placed in acetone for 20 minutes and then in isopropanol for 5 minutes and gently dried using a nitrogen gun, followed by baking at 45°C for 10 minutes on a hotplate (Fig.1b3). The majority (> 50%) of the graphene structures survived the lithography and subsequent drying processes and an optional critical point drying process did not substantially improve the fabrication yield[12].

**Proof mass release:** In order to release the proof masses and suspend them on the double-layer graphene ribbons, the BOX layer (2 μm SiO$_2$) was partly etched from the backside of the SOI substrate by RIE, followed by vapour HF etch to remove the remaining SiO$_2$ layer (Fig.1b4). This two-step etching process was employed to minimize the risk of damaging the graphene. For etching the BOX layer, the device substrate was attached to a 100 mm diameter silicon carrier wafer and all 4 sides of the device substrate were sealed with a tape. Then RIE etching was employed to etch approximately 1.9 μm of the BOX layer, leaving a 100 nm thick BOX layer that was suspending the silicon proof masses. Vapour HF was then used to etch the remaining BOX layer using a custom-built vapour HF etching setup. We used 25 % of HF in the vapour HF chamber and the substrate temperature was set to 40°C. Vapour HF etching of the 100 nm thick BOX layer typically took 5-10 minutes. During BOX etching the graphene was not exposed to the vapour HF, however after the BOX layer was removed at the end of the etching step, the vapour HF could reach the graphene surface. To limit the resulting exposure time of the double-layer graphene to vapour HF to a few seconds, we carefully timed this etching step. As a result, we did not observe any negative effects of the vapour HF etching on the graphene[56]. Our entire fabrication process is compatible with commercial semiconductor and micro-electromechanical (MEMS) foundry processes and can be implemented once wafer scale graphene transfer evolves from laboratories to fabs[57,58].

**Device packaging**

Once the devices were fabricated, the chips were mounted in a ceramic chip carrier with an open cavity. Gold wire bonding was used to connect the electrode pads on the device substrate to the bond pads of the chip carrier. The layout of the metal contacts is schematically shown in Fig.1a, a SEM image of a wire bonded device is shown in Fig.1c5, a photograph of a die with one single accelerometer device with bond pads is shown in Fig.1c6 and a packaged chip after wire bonding is shown in Fig.1c7.

**Basic characterization of graphene devices**

Optical microscopy, white light interferometry (Wyko NT9300, Veeco), Raman spectrometry (alpha300 R, WITec) and SEM imaging were used to observe and characterize the morphology of the devices during and after device fabrication (Supplementary Section S2). White light interferometry was used to detect $SiO_2$ residues inside the trench structures, and to measure the height of the silicon proof masses in relation to the substrate surface after vapour HF etching of the BOX layer. This was done to confirm that the proof masses are fully released (Supplementary Section S3). The white light interferometry measurements show small static displacements of the released proof masses in relation to the substrate surface, indicating that rounded edges with sub-100 nm radiuses at the trenches may cause some pre-straining of the graphene ribbons (Supplementary Section S3). A probe-station connected to a parameter analyser (Keithley SCS4200, Tektronix) was used for preliminary electrical characterization of the double-layer graphene. Two-point and four-point probe measurements were used for measuring sheet resistance, contact resistance, and current-voltage (I-V) properties of the graphene (Supplementary Section S4).

**Static and dynamic mechanical characterization of graphene devices**

For static mechanical characterization, force versus proof mass displacement of devices 8-11 was measured using indentation with an AFM (Dimension Icon, Bruker) tip at the centre of the suspended proof masses. For these experiments an AFM diamond tip (tip radius = 20 nm) with a silicon nitride cantilever (Olympus AC240TS) was used. For the dynamic mechanical characterization, the device resonance frequency and the amplitude of the proof mass motion was measured using a MSA-500 Micro System Analyser (Polytec), which integrates scanning laser Doppler vibrometry (LDV), stroboscopic video microscopy and scanning white light interferometry. The laser spot size of this tool is on the order of 1 μm. For the measurements, the graphene samples were fixed on a small metal plate and then loaded on the piezo disk of a shaker (LDS V 201) with piezoelectric excitation. A commercial reference accelerometer (PCB M352C65) was mounted on the shaker for real-time detection of the acceleration introduced by the shaker. In these experiments the oscillation amplitude of the suspended proof masses attached to the graphene ribbons was studied while they were subject to a time varying actuation caused by the shaker. The mechanical resonances are distinguished as peaks of large oscillation amplitude that occur when the actuation frequency is swept across the resonance frequency of the graphene resonators.

**Electromechanical measurements of acceleration**

During the electrical measurements, in order to shield the graphene devices from mechanical and electrical noise interferences of the measurement system and the environment, a special shielding box with high-performance and low-frequency electromagnetic shielding was designed. Both the device package and the electronic measurement circuits were encapsulated in this box. Two internal boxes made of a ferromagnetic alloy were used for reducing crosstalk between the graphene devices and the measurement circuits. The air-bearing shaker and the dynamic signal analyser were controlled by a computer interface to apply a defined acceleration and read out the sensing signal in form of the output voltage (Supplementary Section S8). In all experiments we used an acceleration frequency of 160 Hz with a 1 g gravitation bias. The acceleration frequency of 160 Hz was used because on one hand, this frequency is sufficiently high to obtain relatively low 1/f-noise and is sufficiently close to 159.2 Hz, which is a commonly used frequency for accelerometer calibrators and is equivalent to a radian frequency of 1000 rad/s (equivalent to $2 \times \pi \times 159.2$). On the other hand, it is well below the intrinsic resonance frequencies of the spring-mass systems of our devices. In addition, 160 Hz avoids the commonly known 50 Hz noise sources and its multiples. The measurement circuits consisted of a first-order high-pass filter with a cut off frequency of 0.079 Hz, and a preamplifier (LT1001OP, Linear Technology) with an amplification factor of approximately 500. The resistance of the graphene ribbons was used as the input resistance of the first-order high-pass filter, supplied with an adjustable DC current (30-100 μA) (Supplementary Section S8). When the proof mass was displaced, the resistance of the suspended graphene ribbons changed due to the piezoresistivity of graphene. The output voltage induced by the change of the resistance was read by a first-order high-pass filter with a cut off frequency of 0.079 Hz to filter any DC drift observed in the graphene resistors, amplified by the amplifier, read by a first-order high-pass filter with a cut off frequency of 15 Hz and then recorded by a dynamic signal analyser (HP 35670A), and finally recorded through the computer interface. Using the output voltage U, the corresponding resistance change could be directly extracted from the relation $\Delta R = U/I$. The resistance change of a device ($\Delta R$) refers to the resistance change of the two suspended graphene ribbons and the resistance of a device (R) refers to the overall resistance, including the resistances of the suspended graphene ribbons, the resistances of the graphene parts that are placed on the $SiO_2$ surfaces and all contact resistances. Furthermore, we have measured the expected linear relation between the voltage output in dependence of the measurement current of our graphene devices (Supplementary Section S13). Before and after each measurement, a multimeter was used to measure the resistance of the graphene ribbons, in order to confirm that the suspended graphene ribbons with the attached proof mass were intact before and after the measurements. In addition, optical microscopy and SEM imaging were used to confirm the mechanical integrity of the devices after the measurements.

**Data availability**

The data that support the plots within this paper and other findings of this study are available from the corresponding author upon reasonable request.

**Code availability**

High-level description of the FEA model of the devices is available from the corresponding author upon reasonable request.

NEMS devices. *Microelectron. Eng.* **159,** 108–113 (2016).

## Acknowledgements

This work was supported by the European Research Council through the Starting Grant M&M's (No. 277879) and InteGraDe (307311), the Swedish Research Council (GEMS, 2015-05112), the China Scholarship Council (CSC) through a scholarship grant, the German Federal Ministry for Education and Research project NanoGraM (BMBF, 03XP0006C), and the German Research Foundation (DFG, LE 2440/1-2). Funding through the European Commission (Graphene Flagship, 785219) is gratefully acknowledged. The authors thank Cecilia Aronsson for help with device processing, Mikael Bergqvist for support with the measurement setup, Matthew Fielden for help with AFM indentation experiments and Jochen Schell for help with LDV experiments. The authors also thank Cristina Rusu, Dimitar Kolev and Pontus Johannisson for discussions about LDV characterization.

## Author contributions

X.F., F.N., F.F., A.C.F., A.D.S., and M.C.L. conceived and designed the experiments. A.D.S., S.W., M.Ö. and M.C.L. developed the graphene transfer method. S.S. performed packaging of all devices. F.F. designed the measurement circuits and contributed in acceleration measurements. S.W. did the Raman characterization. X.F. fabricated the devices (substrate preparation, graphene transfer and patterning, and proof mass release) and performed the experiments, including devices characterization (optical microscopy, SEM imaging, white light interferometry, AFM tip indentation, LDV measurements, and electrical characterization) and acceleration measurements and wrote the manuscript. F.N. provided guidance in the experiments and manuscript writing. X.F., F.F., H.R. F.N., and M.C.L. analysed the experimental results. X.F., H.R. and F.N. analysed the simulation results. All authors discussed the results and commented on the manuscript.

## Competing financial interests

The authors declare no competing interests.

## Additional information

**Supplementary information** is available for this paper at http://doi.org/10.1038/xxxxxx.

**Reprints and permissions information** is available at www.nature.com/reprints.

**Correspondence and requests for materials** should be addressed to F.N. or M.L. or X.F.

**Publisher's note:** Springer Nature remains neutral with regard to jurisdictional claims in published maps and institutional affiliations.

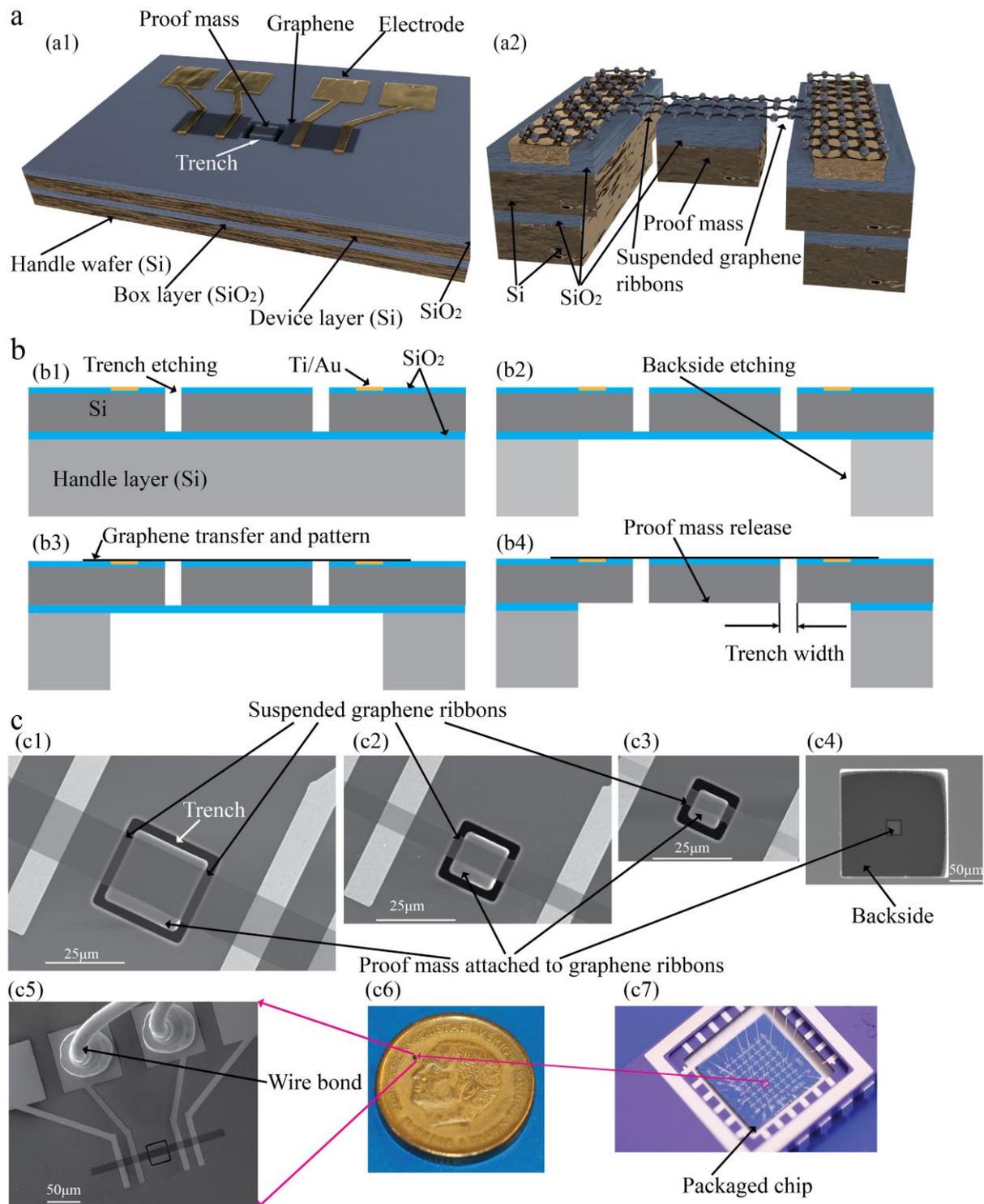

**Figure 1 | Design and fabrication of graphene ribbons with suspended silicon proof masses. a**, Schematic of graphene device. **a1**, Device structure. **a2**, 3D illustration of suspended graphene ribbons with attached proof mass. **b**, Schematic of device fabrication. **b1**, Ti/Au electrodes are embedded into the SiO$_2$ layer and then the trenches are etched in the 16.4 μm thick silicon device layer of the SOI substrate to form the silicon proof masses.

**b2**, The 400 μm thick silicon handle substrate is etched by DRIE in the areas below the proof mass, leaving the silicon proof mass suspended on the BOX layer. **b3**, The double-layer graphene is transferred to the SiO$_2$ surface of the SOI device substrate and patterned by photoresist masking and O$_2$ plasma etching. **b4**, The silicon proof mass is released by sacrificially etching the BOX layer in a 2-step etching sequence, using first RIE etching followed by vapour HF etching. **c**, SEM images and packaging of the devices. **c1-c3,** SEM images with top views of three different devices with identical lengths of the suspended graphene ribbons of 2 x 3 μm, and side lengths of the squared masses of 25 μm, 15μm and 10 μm, respectively. **c4**, SEM image of the backside of the device shown in **c6**. **c5**, SEM image of a device with bond wires. **c6,** Die containing a single device with bond pads, placed on a coin. **c7**, Packaged and wire bonded die containing 64 devices.

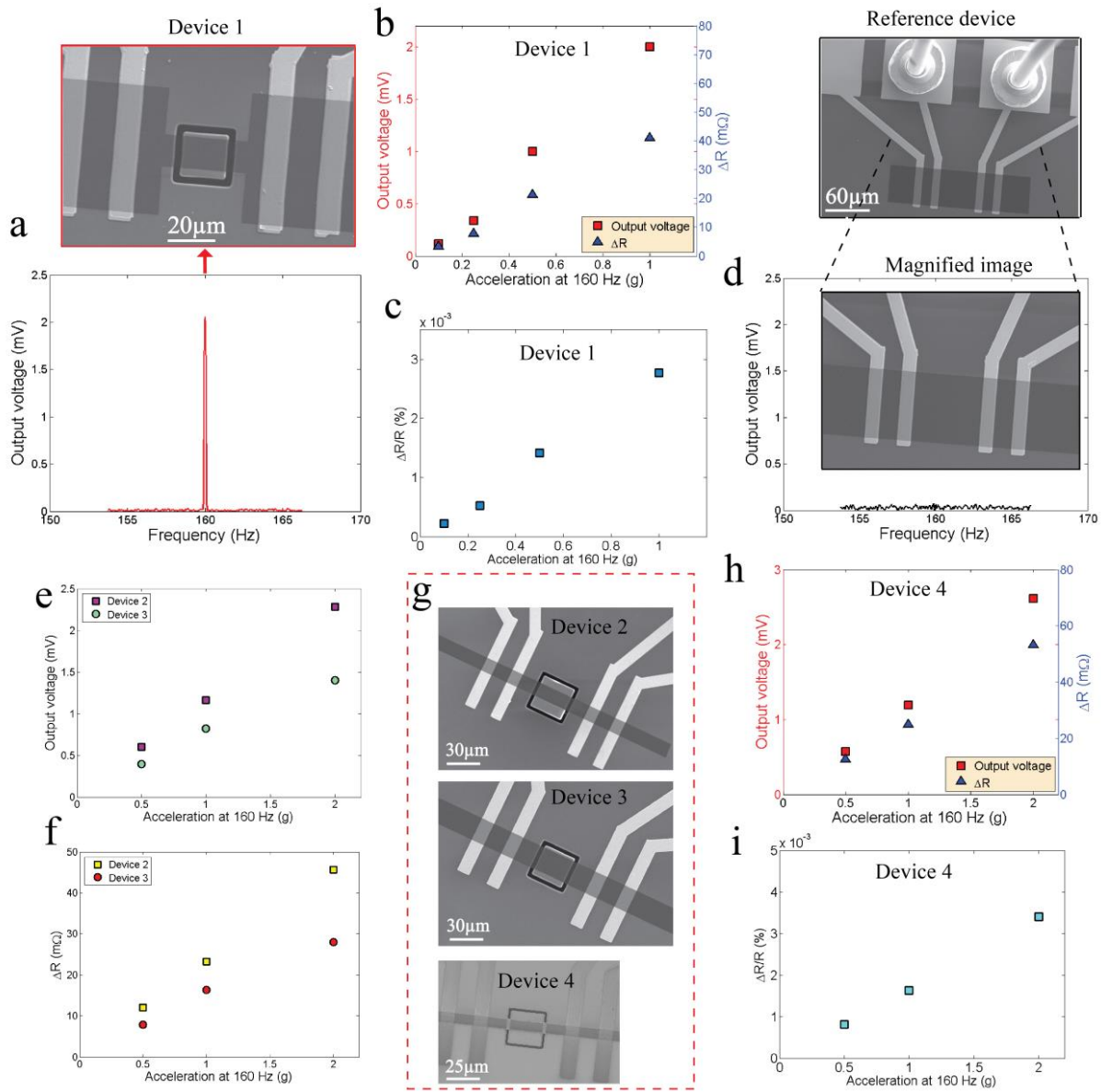

**Figure 2 | Electromechanical characterization of suspended graphene ribbons with attached proof masses. a,** Measured spectrum of the output voltage of device 1 for an acceleration of 1 g at a frequency of 160 Hz as measured by the reference accelerometer of the shaker. For device 1 the dimensions of the proof mass are 20 μm × 20 μm × 16.4 μm, the graphene ribbons are 2 × 3 μm long and 13.5 μm wide. **b,** Measured output voltage and absolute resistance change of device 1 for different accelerations at a frequency of 160 Hz as measured by the reference accelerometer of the shaker. **c,** Measured relative resistance change (ΔR/R) of device 1 for different accelerations at a frequency of 160 Hz as measured by the reference accelerometer of the shaker. **d,** Measured spectrum of the output voltage of the

reference device for an acceleration of 1 g at a frequency of 160 Hz as measured by the reference accelerometer of the shaker. In device 1, the acceleration causes a resistance change in the graphene ribbons, which is not observed in the reference device. **e** and **f,** Comparison of the output voltages and resistance changes of devices 2 and 3 for different accelerations at a frequency of 160 Hz as measured by the reference accelerometer of the shaker. **g,** SEM images of devices 2 to 4. Devices 2 and 3 have identical proof mass dimensions of 30 μm × 30 μm × 16.4 μm and identical lengths of the suspended graphene ribbons (2 × 3 μm), but different ribbon widths (13 μm in device 2 and 24 μm in device 3). **h,** Measured output voltage and absolute resistance change of device 4 for different accelerations at a frequency of 160 Hz as measured by the reference accelerometer of the shaker. For device 4 the dimensions of the proof mass are 25 μm × 25 μm × 16.4 μm, the ribbons are 2 × 2 μm long and 10 μm wide. The centre line of the proof mass is located ~ 2 μm beside the centre line of the graphene ribbons. **i,** Measured relative resistance change (ΔR/R) of device 4 for different accelerations at a frequency of 160 Hz as measured by the reference accelerometer of the shaker.

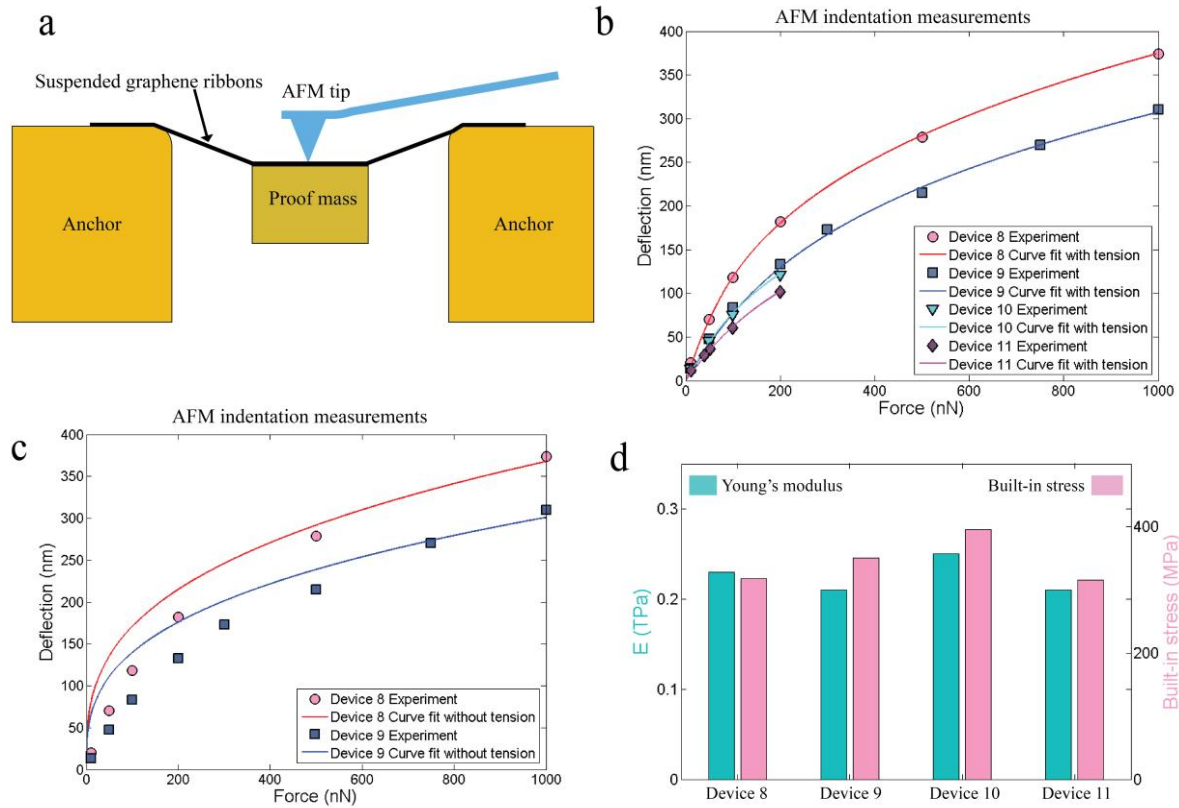

**Figure 3 | Static mechanical characterization of suspended graphene ribbons with attached proof masses. a,** Schematic of the force-displacement measurements by AFM indentation on the centre area of the suspended proof mass. **b**, Force-displacement measurements using AFM indentation of devices 8-11. Device 8, filled circles; device 9, filled squares; device 10, filled triangles (down); device 11, filled diamonds. The length of the suspended ribbons (2 × 4 μm) and the proof mass dimensions (40 μm × 40 μm × 16.4 μm) of devices 8 and 9 are identical. At the same indentation force, for device 8 with a ribbon width of 6 μm, the displacement of the proof mass is larger than for device 9 with a ribbon width of 10 μm, consistent with theory. The length of the suspended ribbons (2 × 3μm) and the proof mass dimensions (25 μm × 25 μm × 16.4 μm) of devices 10 and 11 are identical. At the same indentation force, for device 10 with a ribbon width of 6 μm the displacement of the proof mass is larger than for device 11 with a ribbon width of 10 μm, consistent with theory. The red, blue, cyan and purple solid lines are curve fittings (Matlab function: nlinfit) to the measured AFM force-displacement data of devices 8-11 respectively, using equation (1)

including the term for the built-in tension. **c,** The red and blue solid lines are curve fittings (Matlab function: nlinfit) to the measured AFM force-displacement data of devices 8 and 9, respectively, using equation (1) and assuming zero built-in stress. **d,** Young's modulus and built-in stress values of double-layer graphene extracted from devices 8-11 by AFM indentation experiments.

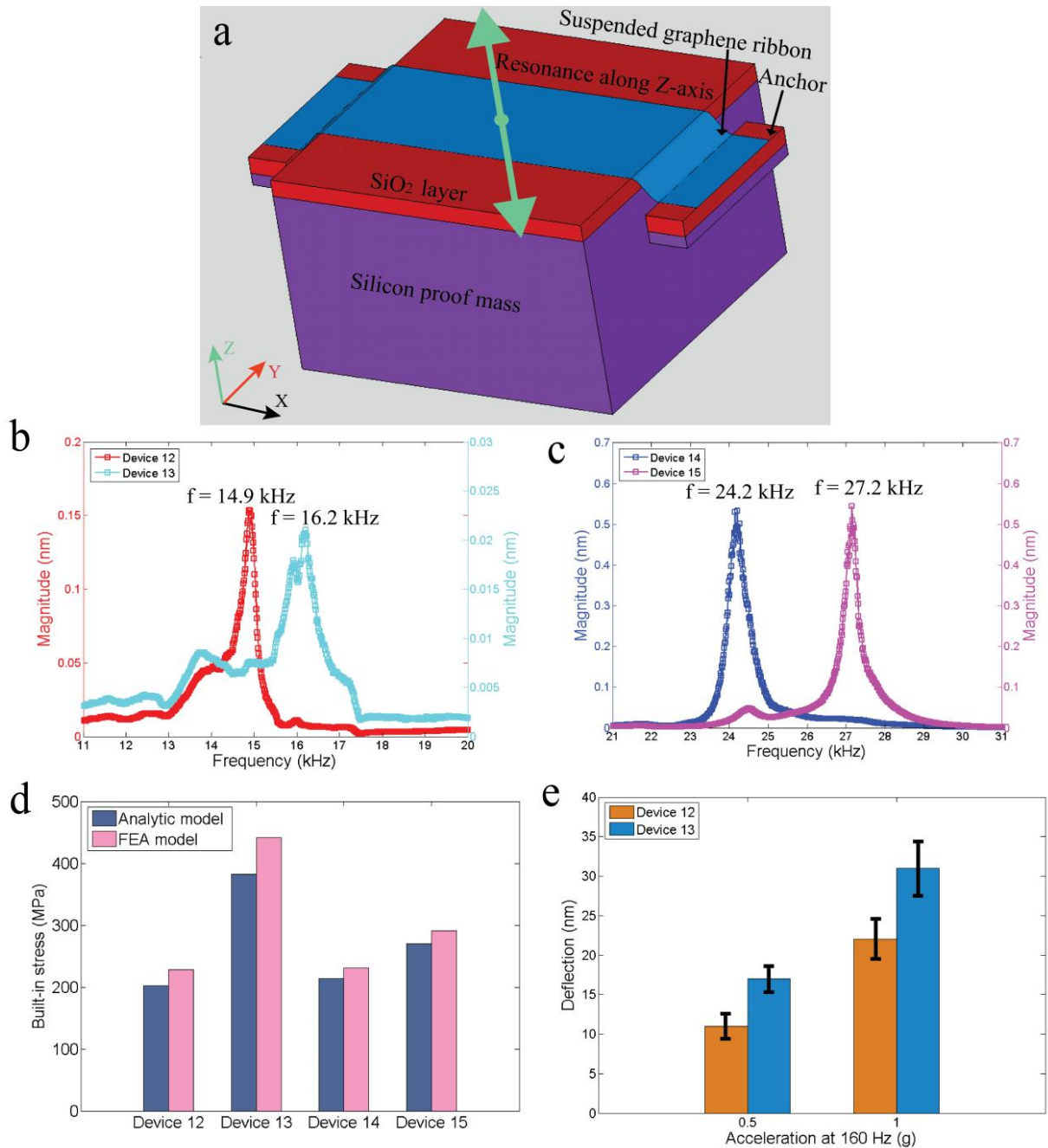

**Figure 4 | Dynamic mechanical characterization of suspended graphene ribbons with attached proof masses. a,** FEA model in Ansys. Three-dimensional model of the device structure, including the different materials (graphene, silicon and $SiO_2$) and the clamping/fixation of the substrate. The coordinate system indicates the axes used in the model. The modal analysis illustrates that the motion in z-axis direction (Z-mode) is the vibration mode of the proof mass that is easiest to excite. **b** and **c,** Measured resonances of devices 12 to 15 using LDV. The resonance frequencies of devices 12 to 15 are 14.9 kHz, 16.2 kHz, 24.2

kHz and 27.2 kHz, respectively. Devices 12 to 15 have identical proof mass dimensions (40 µm × 40 µm × 16.4 µm) and identical length of the ribbons (2 × 4 µm) but different ribbon widths (8 µm for device 12, 5 µm for device 13, 20 µm for device 14, and 20 µm for device 15). It should be noted that the actual acceleration values at the shaker during the frequency sweeps are only of the order of 0.0022 g (device 12), 0.0018 g (device 13), 0.0082 g (device 14), and 0.0032 g (device 15) at frequencies of 14.9 kHz, 16.2 kHz, 24.2 kHz and 27.2 kHz respectively, which are resolved in the frequency domain using fast Fourier transform (FFT). **d,** Comparison of the built-in stresses of devices 12 to 15, extracted by the analytic and FEA models. **e,** LDV measurements of the average displacements of the suspended proof masses of devices 12 and 13 at applied accelerations of 0.5 g and 1 g at a frequency of 160 Hz (see Supplementary Section S19, Table S13). The error bars are the mean deviation of 13 × 13 measurement points.